# Interest-Aware Delivery for Mobile Social Networks: A *TRACE*-driven Approach

Ahmed Helmy {helmy@ufl.edu}
Department of Computer and Information Science and Engineering (CISE)
College of Engineering, University of Florida

[This is a research project proposal]

## 1.Introduction

Future networks will feature applications and services that are highly personalized and human-centric. Social networks applications support and utilize relationships between on-line communities. Mobile networks and portable devices promise to keep users connected anytime anywhere, and can potentially impact their behavior and preferences. At the same time, user mobility and on-line activity impact wireless link characteristics and network performance. Such tight user-device coupling provides a rich set of challenges and opportunities. Fundamental understanding of mobile user behavior is therefore crucial to the design of efficient and effective services for future mobile social networks.

We envision future mobile networks to be human-centric supporting interest-aware delivery, where an interest maybe based on behavior – such as mobility pattern, location, or web browsing - or user profile – such as affiliation, attributes, or activity. An essential capability in future networks will be the ability to provide scalable group communication. Current communication paradigms, including unicast and multicast, require explicit identification of destination nodes (through node IDs or group membership protocols), while directory services translate logical, *interest-specific* queries into destination IDs where parties are then connected using *interest-oblivious* protocols. The power and scalability of such paradigms is quite limited in the context of future, highly dynamic mobile social networks, where it is desirable in many scenarios to support implicit membership based on *interest*. In such scenarios, membership in interest-groups is not explicitly expressed by users, it is rather autonomously inferred by network protocols based on behavioral profiles. This removes the dependence on third parties (via directory lookup), the need for explicit expression of interest (or human intervention), and minimizes delivery overhead to uninterested users. While existing types of social networks will likely evolve in the context of mobile networks, we provide a novel paradigm of communication, deviating significantly from existing approaches and subsequently enabling qualitatively different capabilities for future services in mobile societies. In addition, our services and protocol design philosophy begins by analyzing the context in which the service is to be deployed, unlike most existing wireless networking approaches of designing general purpose protocols.

This proposal introduces a systematic framework, called *TRACE,* for the design and analysis of future interest-aware services in mobile societies. After establishing fundamental understanding and realistic models of mobile user behavior, a trace-driven protocol design is presented. Unlike earlier work on group communication in mobile networks that mainly addresses local areas with tens to hundreds of nodes, our work targets *large-scale* mobile networks with tens of thousands of nodes. Issues of *scalability, efficiency, robustness* and *privacy* in such highly dynamic environment affect the essence of our architectural design and guide our choice of protocol mechanisms. We first introduce our architectural design by discussing the design requirements and presenting an overview of the *TRACE* framework. Then mobile user models and the *profile-cast* protocol are discussed in more detail.

## 2. Service and Protocol Design Requirements

The design requirements are classified into functional (qualitative) requirements, and performance (quantitative) requirements. In terms of functionality, the interest-aware communication service is intended to support – one-to-many - message and query dissemination based on profiles of interest and behavior. The proposed service identifies message receivers implicitly, based on on-line and mobility activity exhibited (and maintained) by the user. Generally, any node can potentially send messages or queries using this service. Delay in delivery maybe tolerated in the infrastructure-less network case, where connectivity can be intermittent. Examples of such service include announcements of seminars or events to specific interest groups on a university campus. In other cases, delivery delay may be critical for emergency alerts to people that frequent (or are affiliated with) certain departments or areas. For those cases notification prioritization based on behavioral patterns, is key to avoid network overload during crises. The service may also be used for discovery of resources with specific behavioral characteristics or mobility capabilities or patterns, as in, for example, message ferries [40]. Such service can lead to automatic configuration and graceful failure recovery for these systems.

In addition, the main factors driving our design are scalability, efficiency, flexibility, robustness and support for privacy.

I. **Scalability**: Unlike related work that considers tens to hundreds of mobile nodes, our architecture should be able to support two to three orders of magnitude more nodes. We believe that mobile nodes will be pervasive, with tens of new classes of application supported by mobile wireless devices (e.g., navigation, location-based services). Several features in our design address scalability: 1. Flat wireless architectures are known not to scale well, mainly due to the far-reaching effects of network dynamics; mobility, failures and topological changes. Such effects consume network resources (i.e., bandwidth, power), and lead to recovery delays and increased route oscillations. Hierarchical architectures, on the other hand, alleviate the above problems, as they tend to localize and dampen network dynamics, and scale routing tables using aggregation. Many existing hierarchical architectures are based on *clustering* mechanisms, in which a cluster-head is chosen to manage each cluster. Such architectures suffer from single point of failure, in which the failure (or movement) of the head may have severe negative effects on the hierarchy. Furthermore, hierarchy maintenance requires dynamic cluster election mechanisms, which usually incur a lot of overhead and complexity. By contrast, we design an architecture that leverages the *small world* structure [3] and behavioral similarity-based clustering [4], eliminating the need for hierarchy maintenance. Such clustering architecture, as we shall show, is inherent in mobile societies, and is established in a fully distributed manner without message exchange. In fact, it is only calculated *on-demand* as needed for message forwarding. 2. Flooding can be quite harmful in wireless environments, as it exhausts network resources, usually incurring unnecessary redundancy and overhead. We avoid the use of flooding, and instead we develop gradient forwarding mechanisms in the *behavioral-space*.

II. **Efficiency**: Future mobile networks will consist of small, portable, energy-constraint devices. For such devices, special attention should be given to energy and storage-constraints. Efficiency metrics considered in our design include: 1. communication overhead, 2. delivery ratio, and 3. delivery delay. The overhead is controlled by the use of gradient-based dissemination and similarity-oriented message forwarding. In addition, no membership maintenance overhead is incurred, as the membership is *implicit*. Efficiency trade-off will also be considered where lower delay maybe achieved with a slight increase in overhead, and so on. The efficiency and trade-offs shall be studied and evaluated via extensive trace-driven simulations and prototype implementations.

III. **Flexibility**: The service should be deployable using various network architectures in multiple contexts: 1. To allow service provisioning in various network architectures we adopt a two-phase approach to our research: *i*. The first phase leads to basic (generic) understanding of mobile user behavior, its representation, structure, stability, and characterization using (in a novel way) data mining and clustering techniques. Such understanding is crucial to the applicability of the interest-aware service in all network architectures.   *ii*. The second phase establishes a new set of distributed protocols that are mainly geared towards mobile peer-to-peer, opportunistic communication schemes. Such environment poses the greatest research challenges, and the solutions therein may then be ported to other network environments as appropriate.

Our *interest-aware communication paradigm* is applicable to several architectures:  *i*. In an infrastructure-based *server architecture*, user profiles may be collected and stored at a directory, then mined for user classification, abnormality detection, or targeted advertisements. *ii*. In a network with low-bandwidth infrastructure, e.g., the *cellular networks*, the behavioral profile exchange and similarity matching may be done over the infrastructure. Then the actual message (data) transfer can occur in peer-to-peer mode. The (perceived) bandwidth limitation may be due to low network capacity, network overload or high cost of transfer. *iii. Decentralized infrastructure-less network* with opportunistic communication. This shall be the focus of the proposed research utilizing stable behavioral profiles for efficient message dissemination. We shall show that developing solutions in such environment produces powerful schemes applicable to other architectures (such as the privacy-preserving scheme).

2. To allow service provisioning in various contexts we provide for two modes of delivery; one where the interest is tightly coupled with behavior and mobility (e.g., music interest maybe inferred from visiting music buildings, libraries or events), and another mode where the interest maybe orthogonal to the monitored behavior (e.g., interest in stamp collection). The former mode utilizes stability of behavioral-profiles for delivery, while the latter utilizes the small world structure in mobile societies.

IV. **Robustness**: In highly dynamic mobile networks, robustness is of prime concern. Robustness is defined as proper operation in the presence of dynamics. Being able to adapt to network dynamics to achieve correct behavior and reasonable performance plays a major role in our design. We conduct behavioral stability analysis and incorporate it into our clustering formation. In addition, our distributed adaptive profile-cast architecture avoids single point of failure scenarios and promises continued operation and graceful recovery during network partitions. We also incorporate path redundancy mechanisms in our *profile-cast* protocol, utilizing the richness of the encounter-graph in mobile societies. Not only does that achieve better performance, but also provides path redundancy that may be used in case of failures.

V. **Privacy**: Exchange of behavioral and mobility profiles may raise issues of user privacy. These concerns are exacerbated in architectures necessitating explicit exchange of behavioral or location traces or the collection of such information at a central location or directory. Several privacy supporting mechanisms drive the design of our architecture: 1. The representation of behavioral-profiles are locally kept at each node and are not exchanged with other nodes. In this sense, *profile-cast* is a non-revealing protocol. This is facilitated via the distributed clustering and similarity-matching schemes, where each node processes delivery and forwarding algorithms locally. 2. The use of Eigen behaviors of association matrices provides only an approximation of the actual behavior, providing a privacy-preserving decision making scheme, further ameliorating privacy concerns.

The *TRACE* framework, discussed next, provides a context-aware systematic approach to protocol design that will enable us to meet the above requirements.

## 3. The *TRACE* Architectural Framework

The *TRACE* framework is introduced to enable *interest-aware* delivery services in mobile networks through behavioral profiling via data monitoring and mining techniques. In addition, we propose to design efficient interest-aware, *profile-cast* protocols for future mobile social networking that heavily draw from the insights and understanding gained through the data mining and analysis.

Our framework provides a systematic method to *T*race, *R*epresent, *A*nalyze, and *C*haracterize mobile societies and contexts, and to *E*mploy context-awareness in modeling and service design, as outlined in Figure 1. In particular, our framework consists of four main components: (1) tracing and behavioral monitoring, (2) data representation and analysis to develop basic understanding of mobile societies, (3) profile characterization through data mining and clustering, and (4) context-aware trace-driven mobility modeling and interest-aware protocol design.

The first component involves collecting data representing network user behavior. In wireless networks such behavior is associated with time, location and duration of login sessions, among other data. To capture such behavior we have established (over the past four years) a library of WLAN traces (*MobiLib*[1]) from major university campuses. We are also collaborating with other researchers that provide complementary data (e.g., Crawdad[21]).

The second and third components are closely related and include data representation, mining and analysis. We plan to extensively analyze individual and collective (group, encounter) user behavior. Our initial analyses of data from four campuses show a highly skewed distribution of location-preference among the user population and show strong similarity of individuals' on-line activities on a daily and weekly basis [2]. Furthermore, our user group analysis shows heavy clustering of users, where a small-world encounter graph forms rapidly - within 16hrs of a 30 day trace- and an individual encounters on average only 2% of the user population [3], in sharp contrast with existing (random and synthetic) mobility models.

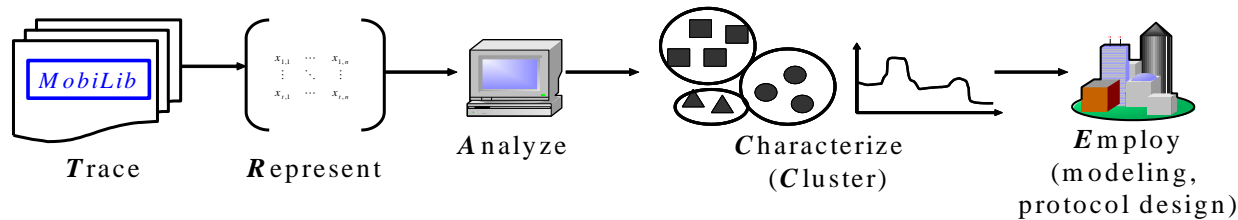

Fig. 1. The *TRACE* framework main components

Such findings are to be employed in the fourth component of our framework to propose a novel realistic mobility model based on time varying communities (*TVC*) [6] and to design the *profile-cast* protocols[5]. Using a user association matrix representation, *profile-cast* utilizes clustering and Eigen vector summary exchange to disseminate messages to nodes with specific profiles of interest and behavior (e.g., based on mobility) [4]. Using an efficient similarity-matching algorithm, we construct a forwarding mechanism for delay tolerant networks (*DTN*s) showing significant promise over existing (epidemic and random walk) *DTN* routing schemes, with initial gains over 50% in reduced overhead and 30% in increased delivery[5].

In the following subsections we shall describe in more detail the various steps of the *TRACE* framework to facilitate and drive the interest-aware service design.

I. *T*race: Our research starts by examining extensive collections of existing libraries of measurements and data sets from the target contexts. The trace collection mainly leverages the ubiquitous wireless network infrastructure on university campuses to record network activities of users, including its time-location information, login and logout events, and the amount of traffic sent and received. With the high adoption rate of WLAN nowadays, this facility provides an extremely rich set of *behavioral* data of users.

This choice of starting point qualitatively distinguishes our approach from the *general purpose* approach of network service design, as illustrated in Fig 2. In previous work, wireless and ad hoc networks designs often started by a *general purpose* protocol design independent of the context in which the protocols will be deployed. This design philosophy led to protocols with limited capability to adapt (and optimize) in different contexts, due to the lack of context-specific information. This approach is especially challenged in the future mobile network context, where the tight user-device coupling and the diversity in *human behavior* in different contexts call for the emergent need of context-aware protocol design. We root the context-aware design in the extensive collection of user behavioral data, the *T*race component.

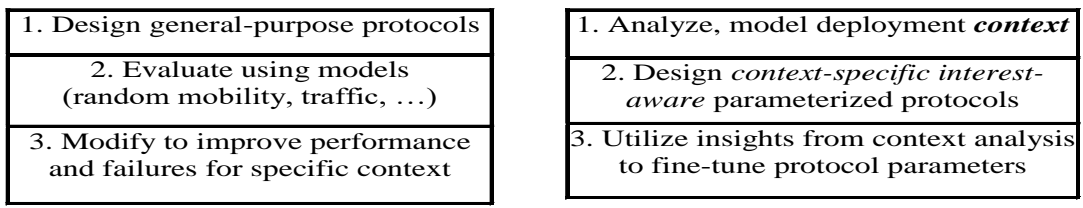

Fig. 2. Contrasting general-purpose design approach to the proposed context-driven approach

We have established a community-wide library of mobile user traces from WLANs of several major university campuses with tens of thousands of users for several years [1]. We also collaborated with and contributed to other community libraries ([21]). It is quite important to investigate traces from multiple sources (instead of just one campus as in most existing work) to be able to alleviate the effect (or bias, if any) of any specific campus or context on the conclusions. By analyzing a rich set of traces from numerous contexts, one would be able to analyze the commonalities and differences, and hence generalize the findings and design to a class or family of contexts. The information should also be augmented (if possible) by contextual information regarding the campus map, building designation and network structure.

Such information is important to have a firsthand understanding of the environment to reduce the imperfection of the trace analysis process (such as ping-pong effects, occasional outages) and its impact on the findings. It is also important for defining the interest and behavioral profiles. Note that the traces are used for analysis purposes only. However, the designed protocol in operation will not assume the availability of the traces in a centralized location. Rather, each node maintains its own trace, as will be described in the protocol design section.

II. *R*epresent and *A*nalyze: The trace is post-processed so that the raw data is transformed into a proper representation to facilitate further analysis. Usually, the raw trace is presented as a sequence of timed events (e.g., association/de-association with specific access points, or session log-in, log-out). The representation of the trace, in our context, is a quantitative measure of these events, presented as a scalar quantity, a vector, a matrix, or a graph. Various representations we choose are based on the specific points we wish to understand about the environment. When the *TRACE* framework is applied to compare multiple traces, this step also involves the normalization of data sets collected with different techniques, so they become comparable [2]. An illustrative example of a matrix representation is given in Fig 3. We shall use that example to illustrate other concepts throughout this document.

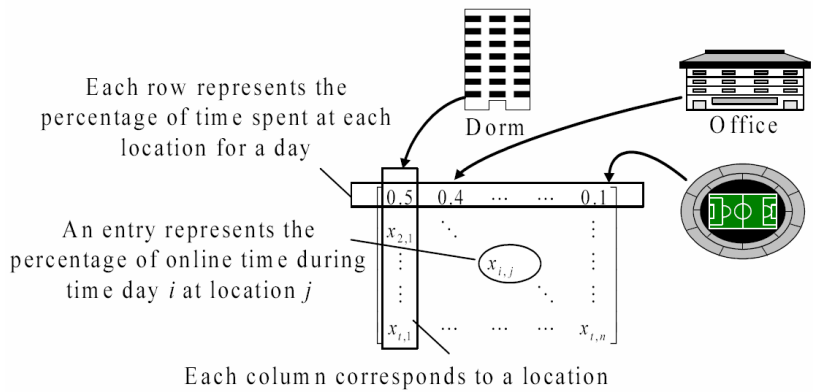

Fig. 3. Example association matrix as behavioral representation of a mobile user

The analysis step involves the application of various mathematical tools and algorithms to obtain distilled information from the representations of the data sets. Examples of such are the distributions for scalar quantities, the singular vectors of the matrices, or the major component of a graph. We plan to investigate various techniques in our work, including basic statistics and probability analysis (leading to distributions of scalar quantities) [2], singular value decomposition (SVD) that reveals the major trends in the matrices distilled from the data sets [4], unsupervised learning (hierarchical clustering), and the application of *small world* theory [3], to enrich the repository of our analysis tools.

III. *C*haracterize: At the heart of the *R*epresentation and *A*nalysis steps is the identification of user behavioral features, or the *C*haracteristics. In dealing with the dynamics in user behavioral data sets, our goal is to use adequate representations and analysis tools to suppress the noise and reveal dominant trends in user behavior. This can be achieved for user mobility preferences, by summarizing a user's location visiting events on each day into an association preference vector, and organizing these vectors in an *association matrix*, as shown in Fig. 3. Principal component analysis (PCA) can then be employed on the matrix representation. We apply singular value decomposition (SVD) to summarize the association matrices. Our initial analysis [4] shows that for most users with a handful of components (i.e., Eigen vectors) with high accuracy (7 of less components capture 90% or more power in the association matrices for most users).

We then analyze the stability of the behavioral principal components for each user over time. The association matrices are constructed based on $d$ days of history, starting from varying time points $T_1$ and $T_2$ with a gap of $T$ days, as shown in Fig 4. The similarity between these components is then calculated for two campuses (USC, Dartmouth) for various values of $d$, as shown in Fig 5. Quite interestingly, we find that this behavioral summary is an intrinsic characteristic for users, as it remains consistent for the duration of several weeks once obtained from a given user (even if based on only 3 or 5 day history). This points to the possibility of using behavioral attributes inferred from the user traces to *identify* the nodes, hence leading to *interest-specific* protocols that deliver messages to a given behavior characteristic, or interest, instead of network identity. Consistent characteristics from multiple data sets reveal the common underlying trends in the mobile social networks. This concept is one of the corner stones of our design.

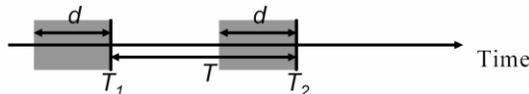

Fig. 4. Consider the trailing $d$ days of behavioral profile at time points that are $T$ days apart

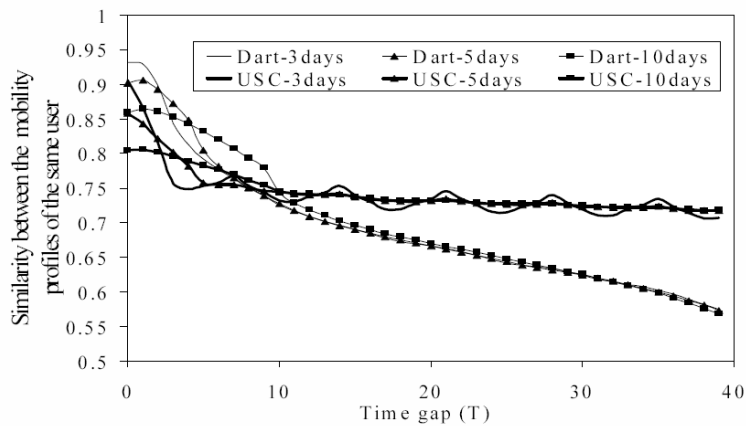

Fig. 5. Similarity metrics for the same user at time gap $T$ apart. Similarity of two sets (i.e., matrices) of Eigen vectors $A$ and $B$ based on their vectors $a_i$'s and $b_j$'s and the corresponding weights $w_{a_i}$'s and $w_{b_j}$'s is given by

$$\sum_{i=1}^{rank(A)} \sum_{j=1}^{rank(B)} w_{a_i} w_{b_j} |a_i \cdot b_j|$$

which is the weighted cosine similarity between the two sets of *Eigen-behavior* vectors.

The use of *user behavioral profiles*, represented by the Eigen vectors of their association matrices (instead of network identities), has several important benefits for protocol and service design in mobile social networks. First, it is an important step towards the design of interest-aware, efficient protocols. For example, based on user behavioral profiles – when two users are *similar*, they are much more likely to encounter[1] and the encounter events last much longer than dissimilar users [41]. This in turn leads to a *small world* structure for the encounter graphs [3]. We leverage this common feature in mobile social networks to guide message transmission decisions. Second, network structure maintenance can now be decentralize and *implicit*, where each node collects and summarizes its own behavior, leading to scalable operation (i.e., overhead in behavior profiling is constant for each node, independent of network size). No explicit construction of clusters is needed; clusters are inherent in the *small world* structure of the mobile societies. Third, the decentralized behavior profiling also helps to protect user privacy, as the behavior characteristics of each user is not known to other nodes. We propose a way to compare user behavioral profiles based on similarity scores that can be calculated locally, removing the need for users to exchange or announce their behavioral characteristics.

---

[1] We say that two mobile nodes *encounter* when they exhibit on-line activity at the same location (or access point) within overlapping time periods. *Encounters* have potential of providing opportunistic peer-to-peer communication in infrastructure-less, intermittently connected, networks.

IV. *E*mploy: The established understanding of mobile society structure is then employed in various tasks, including: 1. building realistic models of mobile societies, (2) classifying users based on behavior and interest, and (3) designing protocols for message and query dissemination.

Realistic models that incorporate user characteristics and the relationship between users are the cornerstones for network protocol analysis. We propose the *time-variant community model* which captures varying mobility characteristics depending on space and time.

For classifying and profiling users, we propose to utilize location visiting preferences to divide the population into distinct groups. We introduce metrics for the *distances* between users in terms of the similarity in location visiting preferences, and leverage unsupervised learning techniques (e.g., hierarchical clustering) to identify the important groups.

Finally, we create a distributed similarity-matching mechanism to design an *interest-aware* message dissemination protocol, called *profile-cast*, when the goal is to deliver messages to a group defined by its behavior or interest. The protocol enables efficient message delivery by using behavioral profiles to guide forwarding decisions. This leads to very high delivery rates while dramatically reducing overhead when compared with *behavior-oblivious* protocols (e.g., variants of epidemic routing or random walk).

In the following two sections we shall elaborate on the mobility modeling and profile-cast protocol design parts for this proposal.

4. **Modeling Mobile Societies:** The Time-variant Community model (*TVC*)

Mobility is one of the key characteristics in future wireless networks. It is, thus, of crucial importance to have suitable mobility models as the foundation for protocol design and analysis. A good mobility model should (i) capture *realistic* mobility patterns of scenarios in which one wants to eventually operate the network; (ii) be *flexible* enough to provide qualitatively and quantitatively different mobility characteristics by changing some parameters of the model, in a repeatable and scalable manner; and (iii) be *mathematically tractable* to allow researchers to derive performance bounds and understand the limitations of various protocols under given sets of scenarios. Most existing mobility models, however, including the *random mobility models* (e.g., random walk, random direction, and random waypoint) or *empirical trace-based mobility models* (e.g., [22],[23],[24]) achieve only parts of the above requirements.

In the time-*variant community mobility model (TVC)* we base our proposal of a *realistic*, *flexible*, and *mathematically tractable* mobility model on extensive analysis of user mobility traces. The model attempts to capture several important mobility characteristics observed empirically from various WLAN traces [2]. The two prominent characteristics are: *i.* the *location preference*, and *ii.* the *time-dependent, periodical behavior* of many nodes. The above important characteristics are included in the design of the TVC model illustrated in Fig. 6 where we construct a periodical schedule of *time periods* to capture omni-present daily and weekly schedules in our daily lives, and thus create a structure in time. We also instantiate popular locations (i.e., the *communities*) for the nodes in the simulation field in each time period, thus creating a structure in space. To our best knowledge, this is the first mobility model that captures non-homogeneous behavior in both *space* and *time*.

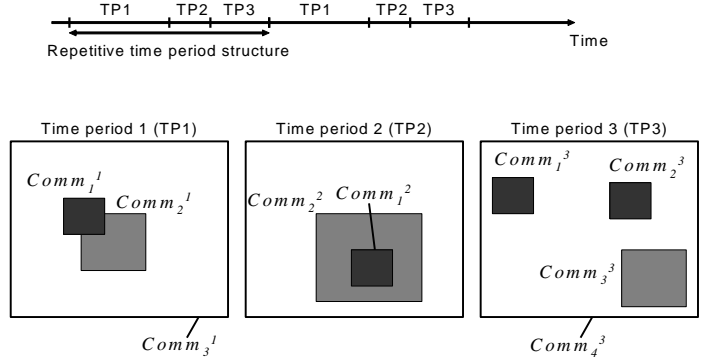

Fig. 6. Illustration of time periods and time-dependent communities in space.

To establish the *flexibility* of our TVC model we attempt to show that we can match its two prominent properties, *location visiting preferences* and *periodical re-appearance*, with multiple WLAN traces collected from environments such as university campuses and corporate buildings. More interestingly, we further compare our TVC model with other qualitatively different traces, and establish that the TVC model is generic enough to have wider applicability. We initially validate this claim by matching our TVC model with two additional mobility traces: a vehicle mobility trace [10] and a human encounter trace[11]. In the latter case, we were able to match our TVC model with other mobility characteristics not explicitly incorporated in our model by its construction, namely the *inter-meeting* time and *encounter duration* distributions between different users/devices. This highlights the potential promise of the TVC model and the driving *TRACE* framework in capturing the decisive factors of typical human mobility. This result substantiates the benefit of our context-aware approach towards network design, and encourages our ongoing and future efforts in that area of research.

In addition to the improved realism, the TVC model can be mathematically treated to derive analytical expressions for important quantities of interest, such as the *nodal spatial distribution*, the *average node degree*, the *hitting time* and the *meeting time* [12]. These quantities are often fundamental to theoretically study issues such as routing performance, capacity, connectivity, etc. To our best knowledge, this is the first synthetic mobility model proposed that matches with traces from multiple scenarios, and has also been theoretically treated to present the above quantities [12]. We also make the initial code of the TVC model available at [13].

We plan to further pursue research on the TVC model to answer questions regarding the clustering and group behavior exhibited in the model, and to further validate the model using other traces from various contexts.

## 5. Interest-aware Mobile Delivery Service: Profile-Cast (*PCast*)

We propose a novel interest-aware communication paradigm to enable a new class of services in mobile societies. Our ability to capture *interest* has thus far been based on *behavior*. Applying such a behavior-driven paradigm in mobile networks poses several research challenges. First, how can user behavior be captured and represented adequately? Second, is user behavior stable enough to enable meaningful prediction of future behavior with a short history? How can such services be provided when the interest or behavior cannot be centrally monitored and processed? And finally, can we design privacy-preserving services in this context? To address these questions we borrow from earlier concepts and findings discussed in the *analysis* and *representation* phases and utilize the stability of the user behavioral profile and implicit structure in the human networks and behavioral space[2] to guide message and query dissemination given a target profile.

*PCast* begins with the *self-monitoring* mechanism, in which each node maintains a history of its *behavioral prof*ile represented using an *association matrix* similar to that in Fig. 3. In the matrix, each row vector describes the percentage of time the user spends at each location on a day, reflecting the importance of the locations to the user. Based on analysis of the traces, such matrix is quite sparse and only a compact representation may be kept. Also, 3 – 5 days of history is usually sufficient to enable meaningful prediction of future behavior. Automatic adaptation of the history window is also possible by performing behavioral stability analysis in the node itself using the similarity-based scheme shown in Section 3.III, Fig 5. For a given user, the singular value decomposition (*SVD*) is applied locally to its association matrix to obtain the Eigen-vectors summarizing the major trends in the matrix. This set of vectors is referred to as the *behavioral profile* of the user. This profile is used in making message forwarding decisions. It can be, however, calculated infrequently (once every few days) by a node then cached and used as needed, minimizing the computation overhead[3].

The initial network architecture considered for *PCast*, and perhaps the most challenging, is that of opportunistic communication with intermittent connectivity (or delay tolerant networks), where peer-to-peer communication is used during *encounter* events. In such architecture message relaying is performed by nodes in a store-carry-forward mode. Our initial analysis and characterization phases show *implicit structure* in mobile human networks, where the similarity in user *behavioral profiles* leads to frequent and long-lasting encounters, providing rich connectivity between clusters of similar users. On the other hand, random encounters occur between dissimilar users providing opportunities for otherwise disconnected groups of nodes to communicate. This provides the basis for the gradient-based forwarding mechanisms for *PCast*.

In *PCast*, a *target profile (TP)* is used to replace network IDs to indicate the intended receiver(s) of a message (i.e., those with matching behavioral profile to the target profile chosen by the sender are the intended receivers). We present two modes of operation under the over-arching paradigm: the *target mode (PCast:T)* and the *dissemination mode (PCast:D)*. The *target mode* is used when the *target profile* is specified in the same context as the *behavioral profile* (i.e., the *target profile* is in terms of *mobility preferences*). The *dissemination mode*, on the other hand, is used when the *target profile* is de-coupled from mobility preferences (e.g., delivering to groups with a specific affiliation or interest, which is independent of the *mobility preference*, as illustrated in Fig. 7).

Upon sending a PCast message, a sender includes one of the following profiles in the header of the message: *i.* if the intended recipients are those users similar to the sender, then the sender includes its own behavioral profile (e.g., cached *SVD*) as the target profile, *ii.* if the intended recipients are similar to another behavioral pattern, the sender constructs the target profile (e.g., considering *SVD* for a *virtual user* visiting the target locations with the required weights) and includes that behavioral profile in the target profile of the message, otherwise *iii.* if the sender cannot map the target interest into a behavioral profile, it indicates a mobility-independent profile (using a flag bit) in the target profile, and includes any interest semantics it wants the recipients to match on to accept the message. This triggers the *disseminate mode* in the PCast protocol. For our initial design we assume that one of the above scenarios (*i,ii* or *iii*) is readily provided by the application using PCast. This shall be further investigated in the proposed research. In cases *i* and *ii* above, the *PCast:T* scheme is used, whereas in case *iii* the *PCast:D* scheme is used, as described below.

---

[2] Based on the user behavior traces, we have established initial empirical validation that the *behavioral profile* of a given user remains highly similar to its former self for durations up to weeks. The surprising observation is that, the similarity metric between a pair of users, based on their respective *behavioral profiles*, predicts their future similarity reasonably well [41]. This demonstrates that the *behavioral profile* we design is an intrinsic characteristic of a given user and a valid representation of the user for a good period of time into the future. We refer to this phenomenon as the *stability* of user *behavioral profiles*, which can be used to map the users into a high dimensional *behavioral space*. The position of users in the behavioral space reflects how similar they are with respect to the *behavioral profile* we construct.

[3] Knowledge of the campus building map (or points of interest or similar) is assumed for the initial design of the protocol (which can be obtained via simple query to neighbors for newcomers). Alternatively, we shall investigate in the proposed research encounter-based virtual maps that are constructed incrementally as a node discovers its immediate neighbors or environment, in which case no prior knowledge of maps is needed.

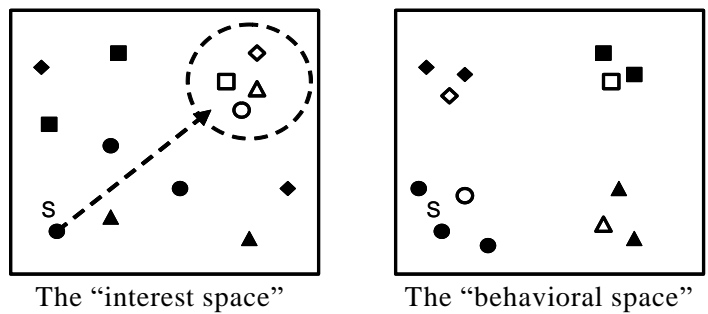

The "interest space"   The "behavioral space"

Fig. 7. Illustrations of the *PCast:D* scheme. Left chart: The goal is to send a message to a group of nodes with a similar characteristic in the *interest space* (white nodes in the circle). Right chart: however, they may not be similar to each other in the *behavioral space* (nodes with the same legend represent similar nodes in the behavioral space).

In *PCast:T* scheme, to deliver messages to receivers defined by a given *target profile*, we gradually move the message towards nodes with increasing similarity to the *TP*, in the hope that such transmissions will improve the probability of encountering the intended receivers. When the message reaches a node close to the *TP* in the behavioral space, most nodes encounter frequently with this node are also similar to *TP*. Hence, the message should be spread to other nodes in the neighborhood in the *behavioral space*. Note that the profile may match users in multiple different locations that have similar interests.

*PCast:T* uses a gradient-ascend approach, in which a forwarding node (initially the sender) during an encounter with another node, sends only the behavioral profile of the message (by sending the corresponding Eigen-vectors, included earlier in the message header, and optionally a distance for gradient-ascend if the message has not reached a similar user yet). The other node then performs similarity matching locally (as described in Section 3.III, Fig 5). If the match is above the similarity threshold (or if the match has a distance less than that indicated in the message header) then a reply is sent to the forwarder to forward the message.

In the *PCast:D* scheme, our aim is to have only one message holder among the nodes who are similar in their behavioral profiles (or equivalently, pick only one message holder within a neighborhood in the behavioral space. In Fig. 7, this corresponds to having only one message holder from each group of nodes with the same legend). We add the messages holders carefully to avoid overlaps in the encountered nodes among message holders. This can be achieved by selecting nodes that are very dissimilar in their *behavioral profiles*, using a gradient-descend forwarding scheme.

We plan to compare our *PCast* schemes to optimal schemes that assume global knowledge and evaluate the improvement over other dissemination schemes. Our initial evaluation shows that, for the *PCast:T mode*, comparing with the delay-optimal protocol, our protocol is close in terms of success rate (more than 94%) and has less overhead (less than 84% to the optimal), and the delay is about 40% more. For the *PCast:D mode*, our protocol features lower storage overhead than the delay-optimal protocol with more than 98% success rate under a storage overhead less than 60% of the delay-optimal protocol, while the delay of *PCast:D* is about 32% more than the optimal. These results again illustrate the power of understanding the context for protocol design – by identifying important characteristics in user behavior and their mutual relationship, we are able to incorporate them in the design of efficient, context-aware communication schemes. In our proposed research we will develop a comprehensive evaluation plan to further evaluate performance trade-offs in various environments and contexts.

## 6. Related Work

We propose a novel *profile-cast* communication paradigm and systematic design framework called *TRACE*. We lay the foundation of this work on a solid analysis of empirical user behaviors, enabled by extensive collections of user behavioral traces. Many of them can be found in the archives at [1], [21]. Due to the rapid adoption of wireless communication technologies and devices, we have seen wide-spread interests in analyzing the traces to understand user behavior. The scope of analysis includes WLAN usage and its evolution across time [14], [15], [16], user mobility [2], [17], [18], traffic flow statistics [19], user association patterns [20] and encounter patterns [3], [11]. Our previous work [2],[3] are among others to explore the space of understanding *realistic* user behaviors *empirically* from large-scale data traces. These efforts in collecting traces from wireless networks and understanding user behaviors provide a fertile ground for behavior-aware research in the given context.

There are several prominent examples of utilizing the data sets for context specific study. *Mobility modeling* is a fundamentally important issue, and several works, including ours, focus on using the observed user behavior characteristics to design realistic and practical mobility models [6], [22], [23], [24]. They have shown that most widely-used existing mobility models (mostly random mobility models, e.g., random walk, random waypoint; see [25] for a survey) fail to generate realistic mobility characteristics observed from the traces. Since the underlying mobility model is a decisive factor for protocol performance [26], it is of crucial importance to solidify our understanding to the context and provide the most pertinent mobility model possible. Our previous work on the TVC model [6],[12] and the on-going work to automate its adaptation to a specific context has its own merit in providing a fundamental environment for protocol and service performance analysis.

The identification of user *behavioral profiles* and the *classification* of users is another major line of research. In this domain, we have shown that user mobility preference matrix representation leads to meaningful user clustering [4]. Our effort is related to several other works with focuses on classifying users based on their mobility periodicity [27], time-location information [28], [29], or a combination of mobility statistics and activeness [30]. The classification of users is important to understand the user dynamics in a given environment, and use this information to facilitate important tasks in the studied context, such as behavior-aware advertisements or better network management. Such behavior-aware services are applicable to several network architectures. In the *centralized server-based architecture*, user profiles could be collected and stored at a data repository, and mined for useful information for context-aware services. In the *cellular networks*, the low-bandwidth channel between the users and the infrastructure can be leveraged to exchange behavioral profiles and match users. The *decentralized infrastructure-less networks* present itself as the most challenging scenario, due to the lack of global knowledge, high network dynamics, and stringent resource constraint.

The major application considered in this proposal is to design a message dissemination scheme in decentralized environments. While several previous works exist in the delay tolerant network field, most of them (e.g., [31],[32],[33],[34]) consider one-to-one communication pattern based on network identities. We propose *profile-cast*, a whole new one-to-many communication paradigm targeted at a behavioral group in decentralized environments. The profile-cast paradigm addresses the need of interest-aware communication – the messages are sent to those who are interested in receiving it, determined by the past *behavioral profiles* of the users and the classification based on it. We propose a fully decentralized and efficient approach to meet this goal, unlike some of the previous works assuming existing infrastructure

(e.g., PeopleNet [35] uses specialized geographic zones for queries to meet and messages are delivered to these zones through the infrastructure) or relying on persistent control message exchanges (e.g., the delivery probability in [33],[34]) for each node to learn the structure of the network, even when there is no on-going traffic. From the design point of view, our approach differs from [33],[34] by avoiding such persistent control message exchanges to achieve better power efficiency, an important requirement in decentralized networks.

The spirit of our design is more similar to [31],[32],[36], in which each user learns the structure of the network locally (i.e., by collecting its own behavioral characteristics and self-profiling, therefore positioning itself in the behavioral space and evaluating its similarity to other users) and uses the information for message forwarding decisions. The major differences between this proposal and the previous works [31],[32],[36] are two fold: First, we design a generic profile-cast framework, in which the *target profile* need not be related to the *behavioral profile* based on which the message dissemination decisions are made. Second, we also provide a non-revealing option in our protocol, thus no node has to explicitly reveal its behavioral pattern or interests to others, as opposed to [31],[36].

Robust geographic routing [37],[38] and geocast [39] can also be used when appropriate, however, three fundamental qualitative differences exist with the proposed scheme: a. the target profile in the profile-cast paradigm can be more generic then the geocast zone. It could be specified not only according to the current behavior of the user, but also containing historical behavioral profile collected through time. b. the gradients followed are in the behavioral space (instead of the physical, Euclidean space), and c. support for forwarding in intermittently connected networks is provided.

## 7. Proposed Research Plan

As was explained above, our preliminary work suggests that interest-aware forwarding can be highly efficient and scalable. There are a number of interesting open questions that we would like to address in the proposed research project. We believe that the answers to these questions are important and will have a high impact on the research community, because the analyses, models and protocols that result from this study will form fundamental building blocks for information dissemination in next-generation mobile networks. We outline some of these questions and possible approaches below.

*I. Trace Analysis and Mining:* We want to answer the question of how to capture user interest. Our ability to capture *interest* has thus far been based on *behavior* using mobility and wireless access activity. As part of this proposal we plan to build upon our initial findings and augment our framework by introducing new dimensions of interest clustering, and classification including study major, gender [7] device preference, and contextual information including building and access point tags, points of interest, landmarks, and activity description, among others.

We also want to investigate the generality of our findings by extending our analysis and mining to a new richer datasets from USC, Dartmouth, UFL and UNC. These shall also be used for further evaluation of the mobility models and protocols.

For data representation we plan to study other vector and matrix representations and evaluate their strengths and weaknesses. In addition, we plan to introduce new matrices at various spatio-temporal granularities and scales and plan to analyze the clustering and small world model for those scales.

Finally, we shall use various data visualization tools to further our understanding of behavioral patterns in mobile societies, including animations in Google Earth, dynamic graph representations, among other tools.

*II. Modeling Mobile Societies:* although we have developed an initial TVC model, we have thus far used only *individual* user behavior analysis for the model development. We plan to further investigate the integration of group and collective behavioral analysis to capture small world and clustering structures that exist in the underlying mobile societies. We shall further test whether the notion of community can, indirectly capture such structure. This would involve developing a *wrapper* around the model to augment its awareness of inter-nodal relationships and correlations.

Further evaluation of the model will be conducted extensively with the use of richer sets of traces. The overall goal will be to provide the reference mobility model for future mobile networks. The simulation code for the model shall be developed, enhanced and released to the community.

*III. Protocol Design and Evaluation:* Our plans for protocol design include three general directions, including adaptivity, privacy and applications.

*i. Adaptivity:* We aim to design a fully self-configuring protocol. To achieve that, the protocol shall incorporate several learning mechanisms, including: *a*. adaptation of the history window to provide reasonable future estimates to perform efficient and correct forwarding decisions, *b*. the ability to learn on-the-fly the contextual maps, tags, encounters and other interest-related information, and *c*. using prediction to build a probabilistic framework for message dissemination management: Our findings indicate promise to investigate and utilize prediction techniques [8]. Our protocols shall subsequently incorporate prediction based on a probabilistic framework to reason about message delivery probability, delay and overhead estimates and bounds. We shall extend the matching algorithms to perform longest matches in a multi-dimensional interest space.

*ii. Privacy*: Further evaluation is needed for the privacy-preserving non-revealing mechanisms developed. How well do these mechanisms preserve-privacy, including the use of Eigen-vector behavioral summaries, and localizing computations (thus avoiding the exchange of behavioral profiles)? This shall be tested using simulations and studies focused on extraction of location, time or behavioral information from the given protocol exchanges. Variants of the proposed schemes shall be developed based on these studies to further enhance privacy characteristics.

Another related issue, is that of excessive unsolicited traffic, or spam. We shall investigate using budget-limited message forwarding techniques (such as limiting the number of hops, time, or overall copies in the network) to alleviate spam effects. Other limitation techniques will be investigated include limiting a sender to only sending to users with a similar behavioral profile. Also, *opt-out* techniques may be used to allow users to decline receipt or forwarding of messages. The effect of such opt-out techniques may limit cooperation and hence may affect protocol performance. Early analysis shows that the encounter pattern between nodes in mobile networks is rich enough to sustain up to 40% of nodes opting out before observing a performance degradation [14]. This is subject to further research.

*iii.* Applications: For our initial design we assume that a behavioral profile is readily provided by the application using PCast. This shall be further investigated by developing a mapping between the application requirements and the interest semantics as understood and supported by the PCast protocol.

Applying PCast to emergency alert notification systems (similar to those deployed on university campuses) will be investigated to prioritize the notifications to reduce the overall network load and cut notification delays users more likely to get affected by the event in question. We also plan to apply our protocols for automatic discovery of message ferries [40].

In addition, extensive evaluation of the protocol using a richer set of data sets and traces from various contexts will be provided. We will also develop a comprehensive evaluation plan to further evaluate performance trade-offs, and to conduct comparisons with optimal protocols, geographic services protocols, including geographic routing, geocast, and location-based services).

The feasibility of the proposed service shall be demonstrated using a multi-step implementation plan: *a*. prototypes shall be developed for mobile handheld devices (including Windows based iPaqs, Linux based Nokia N810s, and Apple's iphone/touch), *b*. initial testing in a test bed in the mobile networking laboratory and the computer science department, and *c*. a proof-of-concept deployment within the university to graduate students and faculty.

Coupled with the research is a comprehensive educational plan to integrate the lessons learned, knowledge, and experiments in mobile societies into the curricula of computer science and engineering at the undergraduate and graduate levels at the University of Florida, as well as underrepresented K-12 schools around north-central Florida.

Following is a timed plan for the proposed research:

|  | Year 1 | Year 2 | Year 3 |
|---|---|---|---|
| Trace Analysis & Mining | - Perform Stability Analysis for various data sets<br>- Augment traces with contextual information | - User Classification and Interest Identification<br>- Trace and Classification Visualization | - Alternative matrix and vector representation analysis |
| Modeling Mobile Societies |  | - Capturing group and collective behavior<br>- Evaluation and validation with various traces and contexts |  |
| Protocol & Service Design | - Adaptivity integration for history and map discovery<br>- Comparison with Optimal and Geographic protocols and Trade-off Evaluation | - Study and Evaluation of Privacy, Cooperation, and Spam reduction mechanisms | - Development and Integration of Prediction Protocols<br>- Application to Emergency Alerts and Ferry Discovery |
| Application Evaluation | - Prototype development on mobile devices | - Test bed Evaluation | - Proof-of-concept Implementation and Deployment |
| Education | - Advise Ph.D. thesis research<br>-Incorporate research as case studies and projects in wireless networks courses<br>-Develop projects in cooperation with local K-12 schools | | |

## Results from Prior Projects:

Prof. A. Helmy, has received an NSF CAREER Award, for the project titled "CAREER: Adaptive Architecture for Multicast Service Support in Large-Scale Mobile Ad Hoc Networks: Design and Evaluation Framework", 6/02-5/07. He is also Co-leading the project "Obtaining Highly Dependable Communication Protocols", 10/02-9/06. Under this project a network architecture was developed to support (1) contact-based resource discovery in ad hoc networks, and (2) rendezvous region-based peer to peer networks in ad hoc networks. On-going work also addresses the impact of mobility on protocol performance in ad hoc and sensor networks. In addition, several algorithms have been developed for systematic (semi-automated) generation of stress test scenarios for networking protocols, with emphasis on multicast-based protocols and ad hoc MAC protocols. Publications resulting from this support include [42, 43, 44, 45, 46, 47, 48, 49, 50].

In addition, he has been a Co-lead on "NeTS-NOSS: Data-Centric Active Querying in Sensor Networks ACQUIRE", 9/04-9/08. Under this project algorithms and routing protocols for query optimization in sensor networks have been developed, analyzed, simulated and implemented, including: the ACQUIRE architecture for active querying in sensor networks, Rugged for gradient-based routing in sensor networks. Publications resulting from this support include [2, 3, 8, 14, 51, 52, 53].

References

[1] MobiLib: Community-wide Library of Mobility and Wireless Networks Measurements (Investigating User Behavior in Wireless Environments) URL *http://nile.cise.ufl.edu/MobiLib/*

[2] Wei-jen Hsu and Ahmed Helmy, "On Modeling User Associations in Wireless LAN Traces on University Campuses", *IEEE Int'l Workshop on Wireless Network Measurements( WiNMee)*, April 2006.

[3] Wei-jen Hsu and Ahmed Helmy, "On Nodal Encounter Patterns in Wireless LAN Traces", *IEEE Int'l Workshop on Wireless Network Measurements( WiNMee)*, April 2006.

[4] Wei-jen Hsu, Debojyoti Dutta, and Ahmed Helmy, "Mining Behavioral Groups in Large Wireless LANs", *ACM MOBICOM*, pp. 338-341, September 2007.

[5] Wei-jen Hsu, Debojyoti Dutta, and Ahmed Helmy, "Poster: Profile-Cast: Behavior-Aware Mobile Networking", *ACM MOBICOM*, September 2007. See also extended technical report at http://arxiv.org/abs/cs/0606002

[6] Wei-jen Hsu, Thrasyvoulos Spyropoulos, Konstantinos Psounis, and Ahmed Helmy, "Modeling Time-variant User Mobility in Wireless Mobile Networks", *IEEE INFOCOM*, May 2007.

[7] U. Kumar, N. Yadav, A. Helmy, "Poster: Gender-based feature analysis in Campus-wide WLANs", *ACM MOBICOM*, September 2007.

[8] J. Kim, Y. Du, M. Chen, A. Helmy, "Comparing Mobility and Predictability of VoIP and WLAN Traces", *ACM MOBICOM Crawdad workshop*, September 2007.

[9] S. Tanachaiwiwat, A. Helmy, "Demo: Proof-of-Concept Worm Interactions via Mobile Devices", *ACM MOBICOM Chants workshop*, September 2007.

[10] Cab Spotting, a project that tracks taxi mobility in the San Francisco Bay Area. Trace available at http://cabspotting.org/api.

[11] A. Chaintreau, P. Hui, J. Crowcroft, C. Diot, R. Gass, and J. Scott, "Impact of Human Mobility on the Design of Opportunistic Forwarding Algorithms," In Proceedings of IEEE INFOCOM, Apr. 2006.

[12] W. Hsu, T. Spyropoulos, K. Psounis, and A. Helmy, "Modeling Spatial and Temporal Dependencies of User Mobility in Wireless Mobile Networks," UF CISE technical report REP-2008-445, online copy available at http://www.cise.ufl.edu/submit/ext_ops.php?op=list&type=report&default=1

[13] Simulation and synthetic mobility trace generator codes of the TVC model and its detailed description are available at http://nile.cise.ufl.edu/TVC_model/

[14] D. Tang and M. Baker, "Analysis of a Local-area Wireless Network," In Proceedings of the 6th annual international conference on Mobile computing and networking (MobiCom 2000), Aug. 2000.

[15] D. Kotz and K. Essien, "Analysis of a Campus-wide Wireless Network," In Proceedings of ACM MobiCom, September, 2002.

[16] T. Henderson, D. Kotz and I. Abyzov, "The Changing Usage of a Mature Campus-wide Wireless Network,"" in Proceedings of ACM MobiCom 2004, September 2004.

[17] M. Balazinska and P. Castro, "Characterizing Mobility and Network Usage in a Corporate Wireless Local-Area Network," In Proceedings of MobiSys 2003, pp. 303-316, May 2003.

[18] M. McNett and G. Voelker, "Access and mobility of wireless PDA users," ACM SIGMOBILE Mobile Computing and Communications Review, v.7 n.4, October 2003.

[19] X. Meng, S. Wong, Y. Yuan, and S. Lu, "Characterizing Flows in Large Wireless Data Networks," in Proceedings of ACM MobiCom, September 2004.

[20] M. Papadopouli, H. Shen, and M. Spanakis, "Characterizing the Duration and Association Patterns of Wireless Access in a Campus," 11th European Wireless Conference 2005, Nicosia, Cyprus, April 10-13, 2005.

[21] CRAWDAD: A Community Resource for Archiving Wireless Data At Dartmouth. http://crawdad.cs.dartmouth.edu/index.php.

[22] R. Jain, D. Lelescu, M. Balakrishnan, "Model T: An Empirical Model for User Registration Patterns in a Campus Wireless LAN," In Proceedings of ACM MOBICOM, Aug. 2005.

[23] D. Lelescu, U. C. Kozat, R. Jain, and M. Balakrishnan, "Model T++: An Empirical Joint Space-Time Registration Model," In Proceedings of ACM MOBIHOC, May 2006.

[24] M. Kim, D. Kotz, and S. Kim, "Extracting a mobility model from real user traces," In Proceedings of IEEE INFOCOM, Apr. 2006.

[25] F. Bai, A. Helmy, "A Survey of Mobility Modeling and Analysis in Wireless Adhoc Networks", Book Chapter in "Wireless Ad Hoc and Sensor Networks", Springer, October 2006, ISBN: 978-0-387-25483-8.

[26] F. Bai, N. Sadagopan, and A. Helmy, "The IMPORTANT Framework for Analyzing the Impact of Mobility on Performance of Routing for Ad Hoc Networks", AdHoc Networks Journal - Elsevier, Vol. 1, Issue 4, pp. 383 -- 403, Nov. 2003.

[27] M. Kim and D. Kotz, "Periodic properties of user mobility and access-point popularity," Journal of Personal and Ubiquitous Computing, 11(6), Aug. 2007.

[28] N. Eagle and A. Pentland, "Reality mining: sensing complex social systems," in Journal of Personal and Ubiquitous Computing, vol.10, no. 4, May 2006.

[29] J. Ghosh, M. J. Beal, H. Q. Ngo, and C. Qiao, "On Profiling Mobility and Predicting Locations of Wireless Users," in Proceedings of ACM REALMAN, May 2006.

[30] D. Tang and M. Baker, "Analysis of a Metropolitan-Area Wireless Network," Wireless Networks, vol. 8, no. 2-3, pp. 107 -- 120, Nov. 2004.

[31] J. Leguay, T. Friedman, and V. Conan, "Evaluating Mobility Pattern Space Routing for DTNs," in Proceedings of IEEE INFOCOM, April, 2006.

[32] E. Daly and M. Haahr, "Social Network Analysis for Routing in Disconnected Delay-Tolerant MANETs," In Proceedings of ACM MOBIHOC, Sep. 2007.

[33] M. Thomas, A. Gupta, and S. Keshav, "Group Based Routing in Disconnected Ad Hoc Networks", in Proceedings of 13th Annual IEEE International Conference on High Performance Computing, Dec. 2006.

[34] A. Lindgren, A. Doria, and O.Schelen, "Probabilistic Routing in Intermittently Connected Networks," Lecture Notes in Computer Science, vol. 3126, pp. 239-254, Sep. 2004.

[35] M. Motani, V. Srinivasan, and P. Nuggehalli, "PeopleNet: Engineering A Wireless Virtual Social Network." in Proceedings of MOBICOM 2005, Sep. 2005.

[36] P. Costa, C. Mascolo, M. Musolesi, and G. Picco, "Socially-aware Routing for Publish-Subscribe in Delay-tolerant Mobile Ad Hoc Networks," to appear in IEEE Journal on Selected Area of Communications.
[37] M. Zuniga, K. Seada, B. Krishnamachari, and A. Helmy. "Efficient Geographic Protocols for Lossy Wireless Networks". t*o appear in ACM Transactions in Sensor Networks.*

[38] Karim Seada, Ahmed Helmy, and Ramesh Govindan. "Modeling and Analyzing the Correctness of Geographic Face Routing under Realistic Conditions". *Elsevier Ad Hoc Networks Journal special issue on Recent Advances in Wireless Sensor Networks, Vol. 5, no. 6,* pp. 855-871, Aug. 2007.

[39] K. Seada, A. Helmy, "Efficient and Robust Geocasting Protocols for Sensor Networks", *Journal of Computer Communications - Elsevier, Special Issue on Dependable Wireless Sensor Networks*, Vol. 29, no. 2, pp. 151-161, Jan. 2006.

[40] W. Zhao, M. Ammar, and E. Zegura, "A Message Ferrying Approach for Data Delivery in Sparse Mobile Ad Hoc Networks," in Proceedings of ACM Mobihoc 2004, May 2004.

[41] W. Hsu, D. Dutta, and A. Helmy, "CSI: A Paradigm for Behavior-oriented Delivery Services in Mobile Human Networks," UF CISE technical report REP-2008-446, online copy available at http://www.cise.ufl.edu/submit/ext_ops.php?op=list&type=report&by_tag=REP-2008 46&display_level=full

[42] F Bai, N. Sadagopan, A. Helmy, "The IMPORTANT Framework for Analyzing the Impact of Mobility on Performance of Routing for Ad Hoc Networks", AdHoc Networks Journal - Elsevier Science, Vol. 1, Issue 4, pp. 383 - 403, November 2003. Conference version presented in INFOCOM 2003.

[43] F. Bai, N. Sadagopan, B. Krishnamachari, A. Helmy, "Modeling Path Duration Distributions in MANETs and their Impact on Routing Performance", IEEE Journal on Selected Areas in Communications (JSAC), Special Issue on Quality of Service in Variable Topology Networks, To Appear 3rd-4th Quarter 2004. Conference version presented in MobiHoc 2003.

[44] A. Helmy, "Efficient Resource Discovery in Wireless AdHoc Networks: Contacts Do Help", Book Chapter in upcoming book on Resource Management in Wireless Networking by Kluwer Academic Publishers, To Appear May 2004 (Book in Editing).

[45] A. Helmy, "TRANSFER: Transactions Routing for Ad-hoc Networks with eFficient EneRgy", IEEE GLOBECOM, December 2003.

[46] A. Helmy, "Mobility-Assisted Resolution of Queries in Large-Scale Mobile Sensor Networks (MARQ)", Computer Networks Journal - Elsevier Science, Special issue on Wireless Sensor Networks, Vol. 43, Issue 4, pp. 437-458, November 2003.

[47] F. Bai, N. Sadagopan, A. Helmy, "BRICS: A Building-block approach for analyzing RoutIng protoCols in Ad Hoc Networks - A Case Study of Reactive Routing Protocols", IEEE International Conference on Communications (ICC), June 2004.

[48] A. Helmy, S. Garg, P. Pamu, N. Nahata, "CARD: A Contact-based Architecture for Resource Discovery in Ad Hoc Networks", ACM Baltzer Mobile Networks and Applications (MONET) Journal, Kluwer publications, Special issue on Algorithmic Solutions for Wireless, Mobile, Ad Hoc and Sensor Networks, 2005. Conference version presented in IEEE IPDPS/WMAN 2003.

[49] A. Helmy, "CAPTURE: location-free Contact-Assisted Power-efficienT qUery REsolution for Sensor Networks", ACM Mobile Computer and Communications Review (MC2R), Special issue on Wireless PAN and Sensor Networks, Volume 8, Issue 1, pp. 27-47, January 2004.

[50] A. Helmy, "Small Worlds in Wireless Networks", IEEE Communications Letters, pp. 490-492, Vol. 7, No. 10, October 2003.

[51] J. Faruque, A. Helmy, "RUGGED: RoUting on finGerprint Gradients in sEnsor Networks", IEEE International Conference on Pervasive Services (ICPS2004), July 2004.

[52] N. Sadagopan, B. Krishnamachari, and A. Helmy, "The ACQUIRE Mechanism for Efficient Querying in Sensor Networks," First IEEE International Workshop on Sensor Network Protocols and Applications (SNPA'03), May 2003.

[53] N. Sadagopan, B. Krishnamachari, and A. Helmy, "Active Query Forwarding in Sensor Networks (ACQUIRE)," Elsevier Journal on Ad Hoc Networks, 2004.